\documentclass[10pt,twocolumn,amsmath,amssymb,aps,prd,superscriptaddress,secnumarabic]{revtex4-2}

\usepackage{graphicx}
\usepackage{dcolumn}
\usepackage{bm}
\usepackage{hyperref}
\usepackage{mathtools}
\usepackage{array}
\usepackage{tabularx}
\usepackage{multirow}
\usepackage{enumitem}
\usepackage[dvipsnames]{xcolor}

\begin{document}

\title{\textbf{A Case for an Inhomogeneous Einstein-de Sitter Universe} 
}

\author{Peter Raffai}
 \email{Contact author: peter.raffai@ttk.elte.hu}
 \affiliation{Institute of Physics and Astronomy, ELTE E\"otv\"os Lor\'and University, 1117 Budapest, Hungary}
 \affiliation{HUN-REN–ELTE Extragalactic Astrophysics Research Group, 1117 Budapest, Hungary}

\author{Dominika E. R. Kis}
 \affiliation{Institute of Nuclear Techniques, Budapest University of Technology and Economics, 1111 Budapest, Hungary}

\author{D\'avid A. K\"odm\"on}
 \affiliation{Institute of Physics and Astronomy, ELTE E\"otv\"os Lor\'and University, 1117 Budapest, Hungary}

 \author{\\ Adrienn Pataki}
 \affiliation{Institute of Physics and Astronomy, ELTE E\"otv\"os Lor\'and University, 1117 Budapest, Hungary}

\author{Rebeka L. B\"ottger}
 \affiliation{Department of Physics, University of Miami, Coral Gables, FL 33124, USA}
 \affiliation{Center for Astrophysics $|$ Harvard \& Smithsonian, 60 Garden St, Cambridge, MA 02138, USA}
 
\author{Gergely D\'alya}
 \affiliation{L2IT, Laboratoire des 2 Infinis - Toulouse, Universit\'e de Toulouse, CNRS/IN2P3, UPS, F-31062 Toulouse Cedex 9, France}
 \affiliation{HUN-REN–ELTE Extragalactic Astrophysics Research Group, 1117 Budapest, Hungary}


\begin{abstract}
We present a local-to-global cosmological framework in which cosmic acceleration emerges from structure formation in an inhomogeneous Einstein–de Sitter (iEdS) universe, without dark energy. The model exhibits a quasilinear coasting evolution toward an effective Milne state driven by growing inhomogeneities. We test the iEdS model with ${H_0=72.5\ \mathrm{km\ s^{-1}\ Mpc^{-1}}}$ and ${\Omega_{\mathrm{m},0}=0.272}$ using CMB, BAO, and SN Ia data. The iEdS model fits the data comparably to $\Lambda$CDM and resolves the $H_0$ tension, while yielding a cosmic age ${t_0\simeq 13.67\ \mathrm{Gyr}}$ consistent with globular-cluster estimates.
\end{abstract}

\maketitle

\section{Introduction}
\label{sec:intro}

The flat Lambda Cold Dark Matter ($\Lambda$CDM) model is the prevailing cosmological model, describing a spatially flat universe with dark energy in the form of a cosmological constant ($\Lambda$)~\cite{Peebles_Ratra_2003}. It shows remarkable concordance with diverse observations across cosmic epochs~\cite{Weinberg_et_al_2013}, including three standard precision probes: cosmic microwave background (CMB) anisotropies~\cite{Hu_Dodelson_2002,WMAP_2013,Planck_2018}, baryon acoustic oscillations (BAOs) identified in galaxy surveys~\cite{Eisenstein_et_al_2005,Alam_et_al_2021,Adame_et_al_2025}, and the type Ia supernova (SN Ia) distance modulus--redshift relation~\cite{Riess_et_al_1998,Perlmutter_et_al_1999,Brout_et_al_2022b}. Despite its successes, tensions between locally measured parameters and those derived from CMB and BAO observations ---most notably in the Hubble constant ($H_0$) ~\cite{DiValentino_et_al_2021} and the structure growth parameter ($S_8$)~\cite{DiValentino_et_al_2025}--- as well as other anomalies~\cite{Perivolaropoulos_Skara_2022}, may indicate the need for further refinement of this cosmological framework.

The flat $\Lambda$CDM model is based on a global-to-local approach. It assumes large-scale homogeneity and isotropy (the cosmological principle~\cite{Milne_1935}), which imply a global scale factor $a(t)$ evolving with a universal (cosmic) time $t$ according to the Friedmann equations~\cite{Friedmann_1922,Dodelson_2003}. Local inhomogeneities (large-scale structure and gravitational gradients~\cite{Bernardeau_et_al_2002}) and motions (peculiar velocities~\cite{Tsagas_et_al_2025}) are typically treated as perturbations evolving in a uniformly expanding background, or as potential contributors to the global Friedmann dynamics through the nonlinear nature of Einstein’s field equations, an effect known as cosmological backreaction~\cite{Ellis_2011,Buchert_Raskanen_2012}. It has been proposed that backreaction could mimic a cosmological-constant–like term in the global expansion~\cite{Clarkson_et_al_2011}, though the prevailing view, despite numerous counterclaims (e.g.~\cite{Buchert_et_al_2015}), is that its magnitude is too small to account for the observed acceleration of the universe.

In this paper, we present a local-to-global framework showing that the observed cosmic acceleration, well described by an apparent $\Lambda>0$ term, arises naturally from local dynamics in an inhomogeneous Einstein–de Sitter (hereafter {\it iEdS}) universe without invoking dark energy. In Sec.~\ref{sec:Theory}, we show that such an iEdS universe follows a quasilinear coasting evolution, starting from the Einstein–de Sitter state~\cite{Einstein_deSitter_1932} with negligible inhomogeneities at recombination, then transitioning through an era of accelerated expansion, and finally asymptoting to a Milne universe~\cite{Milne_1935} with $a(t)\propto t$ coasting evolution. The acceleration arises as a purely general-relativistic effect of structure formation, which amplifies an effective negative spatial curvature that emerges naturally without breaking spatial flatness. Fitting the iEdS model to the Planck 2018 CMB temperature power spectrum~\cite{Planck_2018}, DESI DR2 BAO data~\cite{DESI_DR2_2025}, and Pantheon+ SNe Ia~\cite{Scolnic_et_al_2022} (Sec.~\ref{sec:Data_fits}), we find it fits the data comparably to $\Lambda$CDM while resolving the Hubble tension. Conclusions are summarized in Sec.~\ref{sec:Conclusion}.

\section{Theoretical foundations}
\label{sec:Theory}

In place of the cosmological principle, we base our framework on two principles:

\begin{enumerate}[label=(\roman*)]
 \item The global evolution of the universe follows the Friedmann equations, with scale factor $a(t)$ and cosmological parameters $\{\theta\}$.
 
 \item This global evolution can be represented by an ensemble of finite-volume regions evolving according to the Friedmann equations, each with its own scale factor $a_i(t)$ and cosmological parameters $\{\theta_i\}$. The global scale factor is then
\begin{equation}\label{eq:a3}
a(t)^3=\frac{\sum_{i} V_{i}}{\sum_{i} V_{0,i}}\equiv \frac{\sum_{i} a_{i}(t)^3 V_{0,i}}{\sum_{i} V_{0,i}},
\end{equation}
where $V_i$ is the volume of the $i$-th region at cosmic time $t$, and $V_{0,i}$ denotes $V_i$ at $t_0$, with $a_i(t_0)\equiv 1$ for all $i$, so that $a(t_0)=1$.
\end{enumerate}

Note that, in principle, each region can have its own cosmic-time analog $t_i$, related to the global time by a lapse function $L_i(t)$ via $\mathrm{d}t_i=L_i(t)\mathrm{d}t$. In most practical cases these functions are indistinguishable from $L_i(t)=1$ and can always be absorbed into the local Hubble parameters $H_i(t)$. Also, $a_i(t)=0$ occurs at the same $t$ for all $i$ if and only if $\{\theta_i\}=\{\theta\}$ for all $i$; otherwise, the time of the Big Bang, defined by $a(t)=0$, corresponds to the moment when all $a_i(t)$ vanish.

The validity of principle (i) is also ensured in standard cosmology by the global symmetries imposed on Einstein’s field equations through the cosmological principle. Principle (ii), while not guaranteed in general, is implicitly assumed whenever cosmological models are tested or parameters are inferred from observations spanning one or more finite volumes. In doing so, however, cosmological probes rely on a third assumption, which we reject:
\begin{enumerate}[label=(\roman*)]
\setcounter{enumi}{2}
 \item For all probed finite volumes, $\{\theta_i\}=\{\theta\}$ (and thus $a_i(t)=a(t)$) holds.
\end{enumerate}
As we show here, attributing accelerated expansion to an additional component (dark energy) distinct from matter, radiation, and curvature is a consequence of principle (iii), and can be avoided when only principles (i) and (ii) are applied. Although cosmic backreaction studies relax principle (iii), they typically assume that spatial curvature within local regions is sourced by their energy densities, rendering deviations of $H_i$ from the global Hubble parameter, $H_i-H$, negligible (see Appendix~\ref{app:sigma_H}). This assumption does not apply to regions evolving according to the Friedmann equations of principle (ii), where the curvature parameters $k_i$ are free and the curvature terms scale as $\propto a_i^{-2}$, independently of local energy densities.

From principles (i) and (ii), the global Hubble parameter follows from differentiating Eq.~\eqref{eq:a3} with respect to $t$ (full derivations are given in a supplementary note in the project repository~\footnote{\label{note:Zenodo_repo}\url{https://zenodo.org/records/18290574}}~\cite{Zenodo_repo}):
\begin{equation}\label{eq:H}
H \equiv \frac{\dot{a}}{a} = \frac{\sum_{i} H_{i} V_{i}}{\sum_{i} V_{i}} \equiv \left\langle H_i \right\rangle,
\end{equation}
with the Hubble constant $H_0 \equiv H(t_0)$. The corresponding global deceleration parameter is
\begin{equation}\label{eq:q}
q \equiv -\frac{\ddot{a}}{a H^2} = \left\langle q_i \right\rangle - 2\frac{\sigma_H^2}{H^2}, 
\end{equation}
where
\begin{eqnarray}
q_i &\equiv& -\frac{\ddot{a}_i}{a_i H^2}, \label{eq:qi} \\
\sigma_H^2 &\equiv& \left\langle H_i^2 \right\rangle-H^2. \label{eq:sigma_H}
\end{eqnarray}
In the absence of dark energy ($\langle q_i \rangle \geq 0$), global acceleration ($q<0$) occurs whenever $\left\langle q_i \right\rangle < 2\sigma_H^2 H^{-2}$.

Neglecting radiation, we consider an iEdS universe represented by regions with matter density (including dark matter) $\rho_{\mathrm{m},i}$ and curvature density $\rho_{\mathrm{k},i}$. From principle (ii) it follows that
\begin{equation}\label{eq:local_Friedmann}
H_{i}^2=\frac{8\pi G}{3}\left( \rho_{\mathrm{m},i} + \rho_{\mathrm{k},i} \right) \equiv \left\langle H_i^2 \right\rangle\left( \Omega_{\mathrm{m},i} + \Omega_{\mathrm{k},i} \right),
\end{equation}
where, in the definitions of the local density parameters $\Omega_{\mathrm{m},i}$ and $\Omega_{\mathrm{k},i}$, we normalized $\rho_{\mathrm{m},i}$ and $\rho_{\mathrm{k},i}$ by the ensemble average $\left\langle H_i^2 \right\rangle$ rather than by the individual $H_i^2$. As a result, $\Omega_{\mathrm{m},i}+\Omega_{\mathrm{k},i}=1$ holds only when $H_i^2=\left\langle H_i^2 \right\rangle$, while the independent evolution of the regions remains unaffected. The volume averages of the local matter and curvature density parameters are 
\begin{eqnarray}
\Omega_\mathrm{m}&\equiv& \langle \Omega_{\mathrm{m},i} \rangle
=\frac{\left\langle H_i(t_0)^2 \right\rangle}{\left\langle H_i^2 \right\rangle}\Omega_{\mathrm{m},0} a^{-3}, \label{eq:Om_rhom} \\
\Omega_\mathrm{k}&\equiv& \langle \Omega_{\mathrm{k},i} \rangle
= \frac{\left\langle H_i(t_0)^2 \right\rangle}{\left\langle H_i^2 \right\rangle}\left\langle \Omega_{\mathrm{k},i}(t_0)a_i^{-2} \right\rangle, \label{eq:Ok}
\end{eqnarray}
with present-day values ${\Omega_{\mathrm{m},0} \equiv \langle \Omega_{\mathrm{m},i}(t_0) \rangle}$ and ${\Omega_{\mathrm{k},0} \equiv \langle \Omega_{\mathrm{k},i}(t_0) \rangle}$ (note that the volume average $\langle \cdot \rangle$ of present-day quantities is implicitly evaluated at $t_0$). By construction, $\Omega_\mathrm{m}+\Omega_\mathrm{k}=1$ at all times. 

In Eq.~\eqref{eq:Ok}, the effective scaling $\left\langle \Omega_{\mathrm{k},i}(t_0)a_i^{-2} \right\rangle \propto a^{-2}$ occurs only if $a_i(t)=a(t)$ for all regions. Moreover, Eq.~\eqref{eq:Ok} allows $\Omega_\mathrm{k}$ to become nonzero even if the universe is globally flat at $t_0$ ($\Omega_{\mathrm{k},0}=0$), or at any other reference time. In fact, regions with $\Omega_{\mathrm{k},i}>0$ expand faster with $t$ and gain more weight in volume averaging than those with $\Omega_{\mathrm{k},i}\leq 0$. Consequently, even a universe that starts with $\Omega_\mathrm{k}\leq 0$ ($\Omega_\mathrm{m}\geq 1$) evolves toward $\Omega_\mathrm{k}>0$ and $\Omega_\mathrm{m}<1$, asymptoting to a Milne universe ($\Omega_\mathrm{k}=1$) as its final state (a conceptually similar idea was explored independently in~\cite{Racz_et_al_2017}, although the numerical implementation employed there was subsequently shown to be flawed~\cite{Kaiser_2017,Buchert_2018}).

In the iEdS case, ${\left\langle q_i \right\rangle = \left\langle H_i^2 \right\rangle H^{-2} \left( \Omega_\mathrm{m} /2 \right)}$, and Eq.~\eqref{eq:q} reduces to
\begin{equation}\label{eq:qEdS}
q=\frac{1}{2}\frac{\left\langle H_i^2 \right\rangle}{H^2}\Omega_\mathrm{m}-2\frac{\sigma_H^2}{H^2}.
\end{equation}
We introduce the global Friedmann equations postulated by principle (i) as
\begin{eqnarray}
&H^2 = H^2 \left( \Omega_\mathrm{M} + \Omega_\mathrm{K} \right), \label{eq:global_Friedmann_I}\\
&q = \frac{1}{2}\Omega_\mathrm{M}+\frac{1}{2}\left( 1+3w_\mathrm{K} \right)\Omega_\mathrm{K}, \label{eq:global_Friedmann_II}
\end{eqnarray}
implying ${\Omega_\mathrm{M}+\Omega_\mathrm{K}=1}$. The corresponding global density parameters are
\begin{eqnarray}
&\Omega_\mathrm{M} \equiv \frac{H_0^2}{H^2}\Omega_{\mathrm{M},0} a^{-3}, \label{eq:OmegaM}\\
&\Omega_\mathrm{K} \equiv \frac{H_0^2}{H^2}\left( 1-\Omega_{\mathrm{M},0} \right) \exp\left[ 3\int_{a}^{1}\frac{1+w_\mathrm{K}(a')}{a'} \mathrm{d}a'\right]. \label{eq:OmegaK}
\end{eqnarray}
Comparing with Eq.~\eqref{eq:qEdS}, we identify
\begin{eqnarray}
&\Omega_\mathrm{M} = \frac{\left\langle H_i^2 \right\rangle}{H^2}\Omega_\mathrm{m} = \frac{\left\langle H_i(t_0)^2 \right\rangle}{H^2}\Omega_{\mathrm{m},0} a^{-3}, \label{eq:OmegaM_II}\\
&w_\mathrm{K}=-\frac{1}{3}-\frac{4}{3}\frac{\sigma_H^2}{H^2}\frac{1}{\Omega_\mathrm{K}}. \label{eq:wK}
\end{eqnarray}

The ratio $\sigma_H^2/H^2$ (and thus $\left\langle H_i^2 \right\rangle/H^2$) in Eq.~\eqref{eq:qEdS} is determined by matter and curvature fluctuations, and ultimately by the initial conditions of inhomogeneities (e.g. inflationary perturbations constrained by the CMB~\cite{Akrami_et_al_2020}; see Appendix~\ref{app:sigma_H}). Deriving its evolution from first principles would require detailed relativistic simulations of structure formation, which is beyond the scope of this work. We therefore derive minimal analytic forms for $w_\mathrm{K}(a)$ and $\rho_\mathrm{K}(a)\equiv 3H^2\Omega_\mathrm{K}/(8\pi G)$ to test whether a physically plausible iEdS evolution can reproduce current observations without dark energy. Within these limitations, the iEdS model is phenomenological and can be mapped onto an evolving dark energy model with a novel $w(a)=w_\mathrm{K}(a)$ parametrization.

In Refs.~\cite{Raffai_et_al_2025,Kodmon_Raffai_2025,Raffai_et_al_2024}, we showed that the late-time coasting evolution $a(t)\propto t$, characteristic of curvature-dominated expansion, fits various cosmological datasets either comparably or better than $\Lambda$CDM. This implies that once matter domination ends (around $a\simeq 0.7$), $\Omega_\mathrm{K}$ can be effectively described by
\begin{equation}\label{eq:approx1}
\Omega_\mathrm{K} \approx \frac{H_0^2}{H^2} \left( 1-\Omega_{\mathrm{M},0} \right) a^{-2}. 
\end{equation}
In the same regime, we further approximate
\begin{equation}\label{eq:approx2}
\frac{\sigma_H^2}{H^2}\approx r^2, 
\end{equation}
implying ${\left\langle H_i^2 \right\rangle/H^2 \approx const.}$, where $r$ is a dimensionless, time-independent free parameter.

Using Eq.~\eqref{eq:approx1}, the global first Friedmann equation becomes
\begin{equation}\label{eq:global_Fr_eval}
H^2=H_0^2\left[ \Omega_\mathrm{M,0}a^{-3} + \left( 1-\Omega_\mathrm{M,0} \right)a^{-2} \right].
\end{equation}
From Eq.~\eqref{eq:wK}, we obtain
\begin{equation}\label{eq:wK_approx1}
w_\mathrm{K}(a)= -\frac{1}{3}-\frac{4}{3}r^2\left[ \frac{\Omega_{\mathrm{M},0}}{1-\Omega_{\mathrm{M},0}}a^{-1} +1 \right].
\end{equation}
Substituting Eq.~\eqref{eq:wK_approx1} into Eq.~\eqref{eq:OmegaK} gives
\begin{equation}\label{eq:rhoK_approx}
\rho_\mathrm{K}(z)= \rho_\mathrm{K,0}\exp\!\left[ -4r^2 a_\mathrm{K}z\right]\left( 1+z \right)^{2(1- 2r^2)},
\end{equation}
where $a_\mathrm{K}=\Omega_{\mathrm{M},0}/(1-\Omega_{\mathrm{M},0})$ and $a=(1+z)^{-1}$ (note that Eqs.~\eqref{eq:wK_approx1}-\eqref{eq:rhoK_approx} recover $\Lambda$CDM for $a_\mathrm{K}=0$ and $r^2=0.5$). In data fits, we use $w_\mathrm{K}(a)$ and $\rho_\mathrm{K}(z)$ directly, without the approximations in Eqs.~\eqref{eq:approx1}-\eqref{eq:approx2}, which were only applied in their derivations. Equations~\eqref{eq:wK_approx1}-\eqref{eq:rhoK_approx} describe an effective curvature that grows from $\Omega_\mathrm{K}=0$ at early times to $\Omega_{\mathrm{K},0}=1-\Omega_{\mathrm{M},0}$ today. The present-day deceleration parameter is
\begin{equation}\label{eq:q0}
q_0 = \frac{1}{2}\Omega_{\mathrm{M},0}-2r^2.
\end{equation}
In the asymptotic future, $r^2$ will not remain constant but converges to zero, so that ${w_\mathrm{K}\to -1/3}$, ${\rho_\mathrm{K}\propto a^{-2}}$, and ${q\to 0}$, corresponding to curvature-dominated coasting expansion.

Spatial curvature affects angular diameter and luminosity distances not only through the expansion function but also via a geometric factor arising as light from distant sources propagates toward the observer, crossing thin shells at various redshifts. Whereas local curvatures enter the expansion through volume averages (Eq.~\eqref{eq:Ok}), this factor requires surface averages over the projected cross-sections of the regions:
\begin{equation}\label{eq:Ok_geom_avg}
\Omega_\mathrm{k}^\mathrm{geom}=\frac{8\pi G}{3}\frac{\{ \rho_{\mathrm{k},i} \}}{\{ H_i^2 \}},
\end{equation}
where $\{X\}=\sum_{i} XA_i/\sum_{i}A_i$ denotes an average over projected areas $A_i=a_i^2A_{0,i}$. Since $\rho_{\mathrm{k},i}\propto a_i^{-2}$, $\Omega_\mathrm{k}^\mathrm{geom}$ remains zero if initially zero. As a projected surface average, however, it is subject to ensemble properties and cosmic variance. Consequently, CMB and Pantheon+ SNe Ia (full-sky) should yield consistent values, whereas DESI DR2 (covering roughly one-third of the sky) could, in principle, differ. Nevertheless, in all cases $\Omega_\mathrm{k}^\mathrm{geom}$ is expected to be negligible, and we therefore fix it to zero in data fits. Our choice is further supported by the negligible impact $\Omega_\mathrm{k}^\mathrm{geom}$ has on BAO and SN Ia fits within the low-redshift ranges of DESI DR2 and Pantheon+ ($z\lesssim2.3$). The only non-negligible impact on data fits occurs for the CMB at the high redshift of recombination ($z_*\simeq 1090$), where a nonzero $\Omega_\mathrm{k}^\mathrm{geom}$ should indeed be taken into account.

\section{Data fits}
\label{sec:Data_fits}

We tested the iEdS and flat $\Lambda$CDM models against the Planck 2018 CMB temperature power spectrum~\cite{Planck_2018} (Sec.~\ref{sec:CMB}), DESI DR2 BAO~\cite{DESI_DR2_2025} (Sec.~\ref{sec:BAO}), and Pantheon+ SNe Ia~\cite{Scolnic_et_al_2022} (Sec.~\ref{sec:SNIa}), imposing the Planck 2018 constraint ${\Omega_{\mathrm{M},0}H_0^2=1431.354\ \mathrm{km^2\ s^{-2}\ Mpc^{-2}}}$ consistently across all probes, following well-established practice~\cite{Planck_2018,Pogosian_et_al_2020}. Our codes and posterior corner plots are available in the public repository~\cite{Zenodo_repo}.

\subsection{Cosmic Microwave Background} 
\label{sec:CMB}

We computed the lensed CMB temperature power spectrum using the \texttt{camb} Boltzmann code~\cite{Lewis_Challinor_2000} with default accuracy settings, implementing the iEdS model as an effective dark-energy cosmology with a time-dependent equation of state given by Eq.~\eqref{eq:wK_approx1}. A full multi-parameter CMB fit is computationally prohibitive, so we fixed all parameters to the Planck 2018 baseline $\Lambda$CDM best-fit values~\cite{Planck_2018}, except for $H_0$ (and thus $\Omega_{\mathrm{M},0}$), and iteratively adjusted $r^2$ at fixed ${H_0=72.5\ \mathrm{km\ s^{-1}\ Mpc^{-1}}}$ to reproduce the best-fit value of $\theta_\mathrm{MC}$, the approximate angular scale of the sound horizon at recombination. We then computed $\chi^2$ using Planck data~\cite{Planck_2018,Planck_2020} over the multipole ranges $l=2$–2508 and $l=2$–29, and performed Anderson–Darling (AD) tests on the standardized residuals to evaluate the $p$-values for normality~\cite{Anderson_Darling_1952,adtest_2024}. 

\begin{figure}
 	\includegraphics[width=\columnwidth]{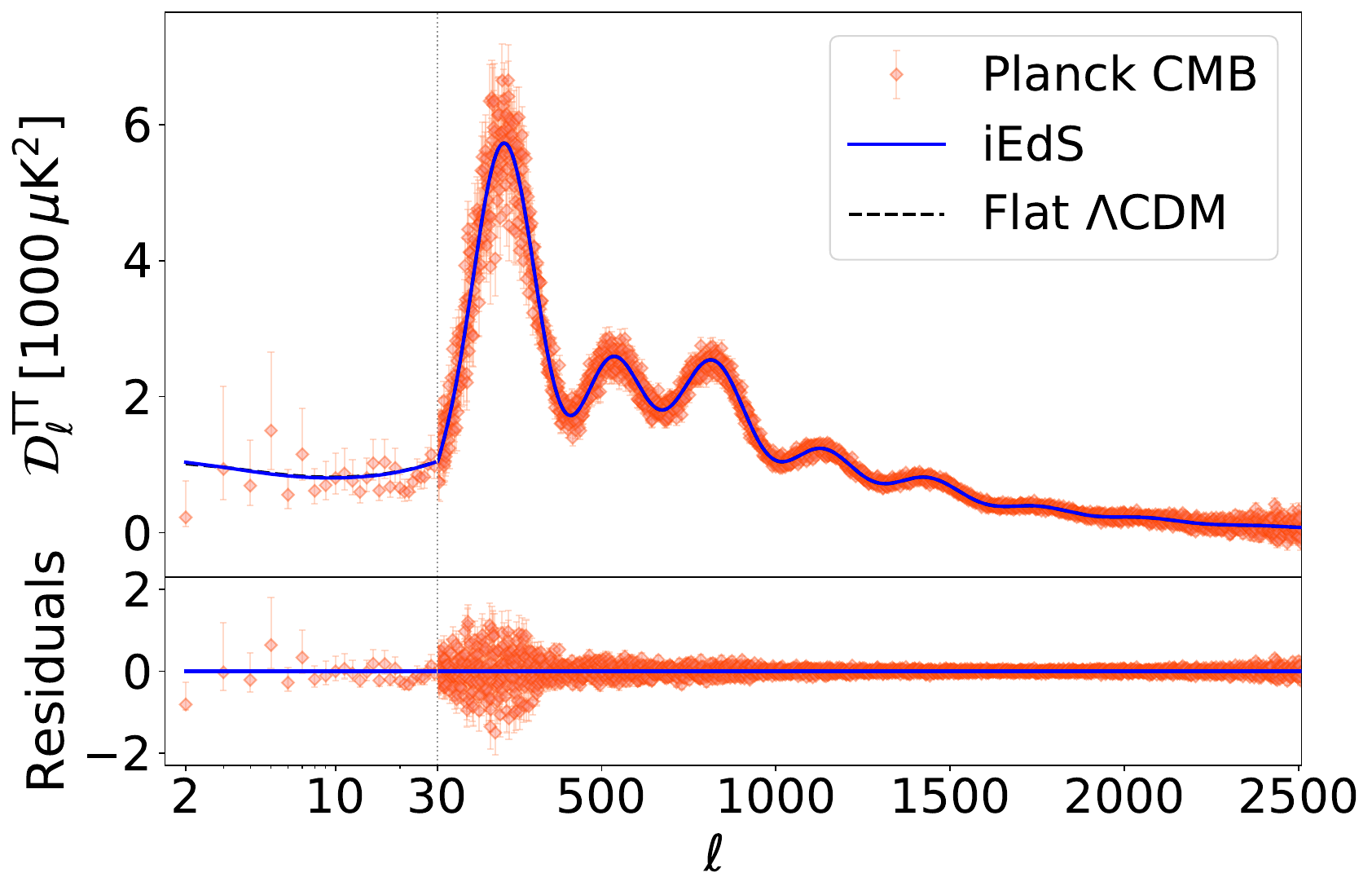}
     \caption{Planck 2018 CMB temperature power spectrum compared with the best-fit $\Lambda$CDM prediction and the iEdS model matched in $\theta_\mathrm{MC}$. The two model spectra are visually indistinguishable, with only minor deviations at the lowest multipoles.}
     \label{fig:fig1}
\end{figure}

The results are summarized in Table~\ref{tab:table1}, and the lensed power spectra of the iEdS and $\Lambda$CDM models are compared in Fig.~\ref{fig:fig1}. For both models and multipole ranges, the residuals pass the AD tests ($p_\mathrm{AD} \ge 0.05$). The iEdS model yields a slightly lower $\chi^2$ value than $\Lambda$CDM, with the spectra visually indistinguishable apart from minor deviations at the lowest multipoles. Using Eq.~\eqref{eq:wK_approx1}, we find $w_\mathrm{K}(1)=-1.09$, close to $w_\Lambda=-1$, while Eqs.~\eqref{eq:approx2} and~\eqref{eq:q0} give ${\sigma_{H,0}^2 /H_0^2 = 0.411}$ and ${q_0=-0.69}$.

\begin{table}
\caption{\label{tab:table1} Model parameters, goodness-of-fit statistics, and AD normality test results for Planck 2018 CMB data. For the iEdS model, ${a_\mathrm{K}=\Omega_{\mathrm{M},0}/(1-\Omega_{\mathrm{M},0})}$, and $\Omega_{\mathrm{M},0}$ is derived from $H_0$ using ${\Omega_{\mathrm{M},0}H_0^2=1431.354\ \mathrm{km^2\ s^{-2}\ Mpc^{-2}}}$.}
\renewcommand{\arraystretch}{1.25}
\begin{ruledtabular}
\begin{tabular}{lcc}
& iEdS & $\Lambda$CDM \\
\colrule
\colrule
$H_0\ [\mathrm{km}\ \mathrm{s}^{-1}\ \mathrm{Mpc}^{-1}]$ & $72.50$ & $67.32$ \\
$\Omega_\mathrm{M,0}$ & $0.272$ & $0.316$ \\
$a_\mathrm{K}$ & $0.374$ & $0$ \\
$r^2$ & $0.411$ & $0.5$ \\
\colrule
$\chi^2\ \left (2\leq l\leq 2508 \right )$ & $2568.6$ & $2570.2$ \\
$p_\mathrm{AD}\ \left (2\leq l\leq 2508 \right )$ & $0.689$ & $0.689$ \\
$\chi^2\ \left (2\leq l\leq 29 \right )$ & $19.657$ & $19.993$ \\
$p_\mathrm{AD}\ \left (2\leq l\leq 29 \right )$ & $0.070$ & $0.053$ \\
\end{tabular}
\end{ruledtabular}
\end{table}

\subsection{Baryon Acoustic Oscillations}
\label{sec:BAO}

We used the \texttt{dynesty} \citep{Speagle_2020} Python package for dynamic nested sampling to jointly fit ${D_\mathrm{M}/r_\mathrm{d}}$ and ${D_\mathrm{H}/r_\mathrm{d}}$ data from DESI DR2~\cite{DESI_DR2_2025}, adopting for both models the Planck 2018 best-fit comoving sound horizon at the drag epoch, ${r_\mathrm{d}=147.049\ \mathrm{Mpc}}$~\cite{Planck_2018}. Here ${D_\mathrm{M}(z)=\int_{0}^{z}c\ \mathrm{d}z'/H(z')}$ denotes the transverse comoving distance (for a flat universe) and ${D_\mathrm{H}(z)=c/H(z)}$ the Hubble distance. The only fitted parameter was $H_0$, with uniform priors ${H_0 \sim \mathcal{U}[63,83]\ \mathrm{km\ s^{-1}\ Mpc^{-1}}}$. The fit minimized
\begin{equation}\label{eq:chi_fit}
\chi^2 = \Delta \bm{D}^{T}C^{-1}\Delta \bm{D},
\end{equation}
where $\Delta \bm{D}$ is the vector of residuals between DESI data and model predictions, and $C$ the DESI covariance matrix. We excluded the LRG1 datapoints, as the $D_\mathrm{M}/r_\mathrm{d}$ measurement at $z=0.51$ shows a $2$–$4\sigma$ tension with both model predictions and Pantheon+ SNe luminosity distances at the same redshift, regardless of the cosmological model (including $w_0w_a$CDM \cite{DESI_DR2_2025}; see also~\cite{Chaudhary_et_al_2025}). After this exclusion, ten BAO points were fitted (results including the LRG1 data points are given in Appendix~\ref{app:BAO_full_fit} to allow comparison). We again performed AD tests for normality on the standardized residuals, now computed as $L^{-1}\Delta \bm{D}$ with $C=LL^T$ the Cholesky factorization of the covariance. Model fits were assessed using the resulting $\chi^2$ values and the Bayes factors ${\log_{10}\mathcal{B}=\log_{10}(Z/Z_\Lambda)}$ from \texttt{dynesty}, where $Z$ is the Bayesian evidence and $Z_\Lambda$ that of $\Lambda$CDM. 

\begin{figure}
 	\includegraphics[width=\columnwidth]{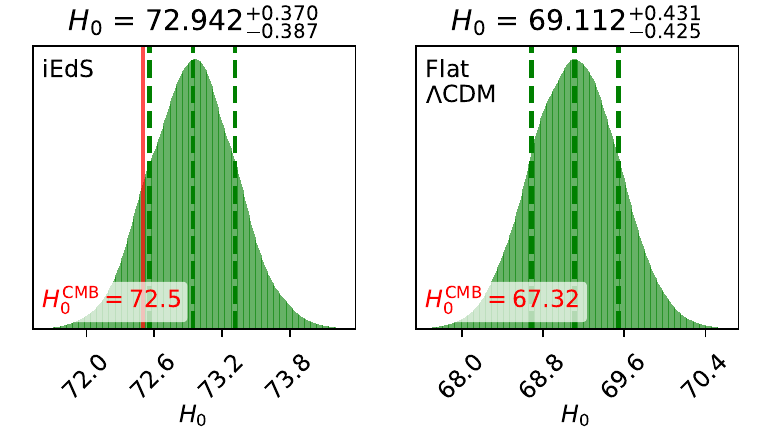}
     \caption{Posterior distributions of $H_0$ from DESI DR2 BAO fits for the iEdS and the flat $\Lambda$CDM models. $H_0$ is in $\mathrm{km\ s^{-1}\ Mpc^{-1}}$. Green dashed lines show the best-fit (posterior median) $H_0$ values with 16th–84th percentile errors listed above each plot, and red vertical lines (where visible) mark the reference $H_0^\mathrm{CMB}$ values (Table~\ref{tab:table1}).}
     \label{fig:fig2}
\end{figure}

\begin{figure}
 	\includegraphics[width=\columnwidth]{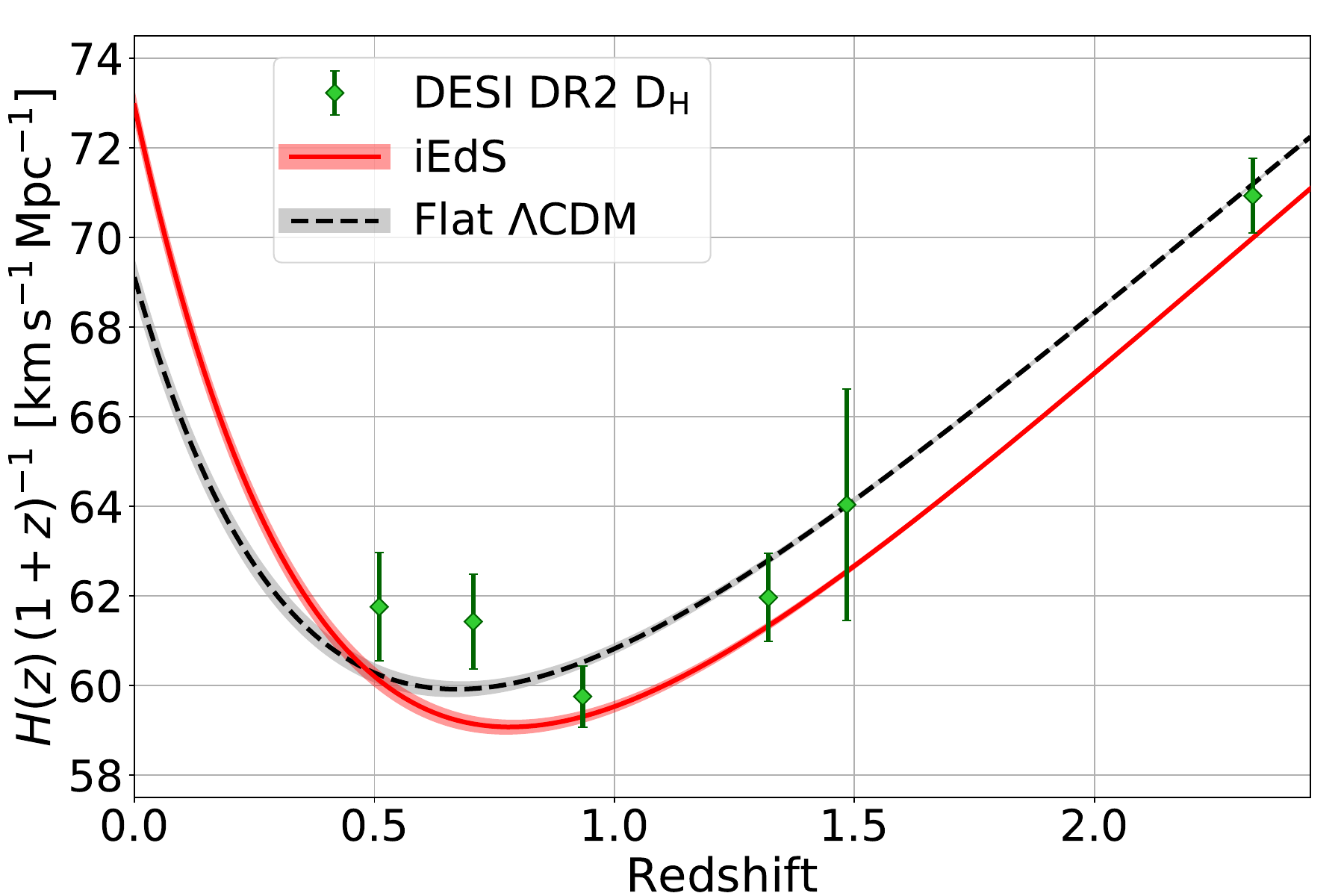}
     \caption{$\dot{a}(z)\equiv H(z)/(1+z)$ from DESI DR2 $D_\mathrm{H}=c/H(z)$ measurements compared with the two best-fit models. The shaded regions show the 16th–84th percentile contours. The LRG1 point at $z=0.51$, excluded from the fits, is shown for visualization only. The transition from decelerated to accelerated expansion occurs at $z_\mathrm{t}=0.782$ and $z_\mathrm{t}=0.672$ for the iEdS and flat $\Lambda$CDM models, respectively.}
     \label{fig:fig3}
\end{figure}

The DESI DR2 BAO data weakly favor $\Lambda$CDM over iEdS, with ${\log_{10}\mathcal{B}=-0.666}$ (Table~\ref{tab:table2}), while both models pass the AD normality test (${p_\mathrm{AD}>0.05}$). Fig.~\ref{fig:fig2} shows the posterior distributions of $H_0$ from the DESI DR2 fits, while Table~\ref{tab:table2} lists the best-fit $H_0$ values (posterior medians with symmetrized 16th–84th percentile errors) and the reference $H_0^\mathrm{CMB}$ from Planck (see Table~\ref{tab:table1}). Table~\ref{tab:table2} also gives the deviations $\left|\Delta H_0^\mathrm{CMB}\right|$ in $\sigma$ units, first considering only fit errors, and in brackets when also including the $\pm0.54\ \mathrm{km\ s^{-1}\ Mpc^{-1}}$ Planck 2018 baseline $\Lambda$CDM uncertainty~\cite{Planck_2018}, which we adopt as a standard for cross-model comparison, although it may differ for the iEdS model. Deviations of $\geq3\sigma$ are marked in red and those of $<3\sigma$ in green. The iEdS model removes the $>4\sigma$ ($>2.5\sigma$) tension seen for $\Lambda$CDM in $\left|\Delta H_0^\mathrm{CMB}\right|$. Fig.~\ref{fig:fig3} shows $\dot{a}(z)\equiv H(z)/(1+z)$ for the two models together with values derived from the DESI DR2 $D_\mathrm{H}$ data. The LRG1 point is included for visualization only, although it was excluded from the fits.

\begin{table*}
\caption{\label{tab:table2} Model comparison results from DESI DR2 BAO and Pantheon+ SNe fits. }
\renewcommand{\arraystretch}{1.25}
\begin{ruledtabular}
\begin{tabular}{llccccccc}
Model & Dataset & $p_\mathrm{AD}$ & $\chi^2$ & $\log_{10}\mathcal{B}$ & $H_0$ & $H_0^\mathrm{CMB}$ & $\left| \Delta H_0^\mathrm{CMB} \right|$ & $\left| \Delta H_0^\mathrm{SN} \right|$ \\
& & $[10^{-2}]$ & & & $[\mathrm{km}\ \mathrm{s}^{-1}\ \mathrm{Mpc}^{-1}]$ & $[\mathrm{km}\ \mathrm{s}^{-1}\ \mathrm{Mpc}^{-1}]$ & $[\sigma]$ & $[\sigma]$ \\
\colrule
\colrule
iEdS & Pantheon+ SNe & $0.058$ & $1565.0$ & $-4.017$ & $72.67\pm 0.99$ & 72.50 & \color{Green} 0.18 \color{black}(\color{Green}0.16\color{black}) &  \\
  & DESI DR2 BAO & $91.74$ & $9.164$ & $-0.666$ & $72.94\pm 0.38$ & 72.50 & \color{Green} 1.14 \color{black}(\color{Green}0.67\color{black}) & \color{Green} 0.25 \\
\colrule
$\Lambda$CDM & Pantheon+ SNe & $0.028$ & $1534.1$ & $0$ & $72.54\pm 0.97$ & 67.32 & \color{red} 5.44 \color{black}(\color{red}4.74\color{black}) &  \\
  & DESI DR2 BAO & $47.6$ & $6.218$ & $0$ & $69.11\pm 0.43$ & 67.32 & \color{red} 4.21 \color{black}(\color{Green}2.61\color{black}) & \color{red} 3.26 \\
\end{tabular}
\end{ruledtabular}
\end{table*}

\subsection{Type Ia Supernovae}
\label{sec:SNIa}

We also tested the iEdS and the flat $\Lambda$CDM models using the Pantheon+ sample of $1550$ SNe Ia ($1701$ light curves at $z\lesssim 2.3$~\cite{Scolnic_et_al_2022})\footnote{\label{note:Pantheon_data}\url{https://github.com/PantheonPlusSH0ES/DataRelease}}. SN distance moduli were computed from the SALT2~\cite{Guy_et_al_2007} light-curve parameters ($m_B$, $x_1$, $c$) following~\cite{Brout_et_al_2022a,Brout_et_al_2022b}:
\begin{equation}\label{eq:dist_mod}
\mu_\mathrm{SN}=m_B+\alpha x_1-\beta c - M_B -\delta_\mathrm{bias} + \gamma \delta_\mathrm{host},
\end{equation}
where $\alpha$, $\beta$, $\gamma$, and $M_B$ are global nuisance parameters, $\delta_\mathrm{bias}$ accounts for selection effects, and $\delta_\mathrm{host}(M_\star)$ corrects for the dependence on host-galaxy stellar mass $M_\star$~\cite{Brout_et_al_2022b,Popovic_et_al_2021,Scolnic_et_al_2022}. Model distance moduli were defined as ${\mu\equiv 5\log_{10}[d_L/10\ \mathrm{pc}]}$, with
\begin{equation}\label{eq:dL}
d_L=(1+z_\mathrm{Hel})\int_{0}^{z_\mathrm{HD}}\frac{c\ \mathrm{d}z}{H(z)},
\end{equation}
where two Pantheon+ redshifts were used: the host-galaxy cosmological redshift in the CMB frame corrected for peculiar velocity ($z_\mathrm{HD}$~\cite{Carr_et_al_2022}), and the heliocentric redshift ($z_\mathrm{Hel}$). Because $M_B$ and $H_0$ are degenerate, we followed~\cite{Brout_et_al_2022b} and replaced $\mu$ by the Cepheid-calibrated $\mu^{\mathrm{Cepheid}}$ from SH0ES~\cite{Riess_et_al_2022} for the $77$ SNe Ia in Cepheid-host galaxies. Keeping the CMB constraint ${\Omega_{\mathrm{M},0}H_0^2=1431.354\ \mathrm{km^2\ s^{-2}\ Mpc^{-2}}}$, we fitted only $H_0$, jointly with the nuisance parameters $\alpha$, $\beta$, $\gamma$, and $M_B$, using \texttt{dynesty} and the $\chi^2$ from Eq.~\eqref{eq:chi_fit}, with $\Delta D_i=\mu_{\mathrm{SN},i}-\mu_i$ (or $\mu_{\mathrm{SN},i}-\mu_i^{\mathrm{Cepheid}}$, where applicable) and $C=C_\mathrm{stat+syst}$ the full covariance from~\cite{Brout_et_al_2022b}. We adopted uniform priors ${H_0 \sim \mathcal{U}[63,83]\ \mathrm{km\ s^{-1}\ Mpc^{-1}}}$, ${M_B \sim \mathcal{U}[-20,-18.8]}$, ${\alpha \sim \mathcal{U}[0,0.2]}$, ${\beta \sim \mathcal{U}[2.5,3.5]}$, and ${\gamma \sim \mathcal{U}[-0.1,0.1]}$ and applied iterative $3\sigma$ outlier rejection until convergence (sigma clipping; see e.g.~\cite{Amanullah_et_al_2010,Riess_et_al_2022}), removing ${N=\left\{ 13,12 \right\}}$ SN data points for the iEdS and flat $\Lambda$CDM models, respectively. 

\begin{figure}
 	\includegraphics[width=\columnwidth]{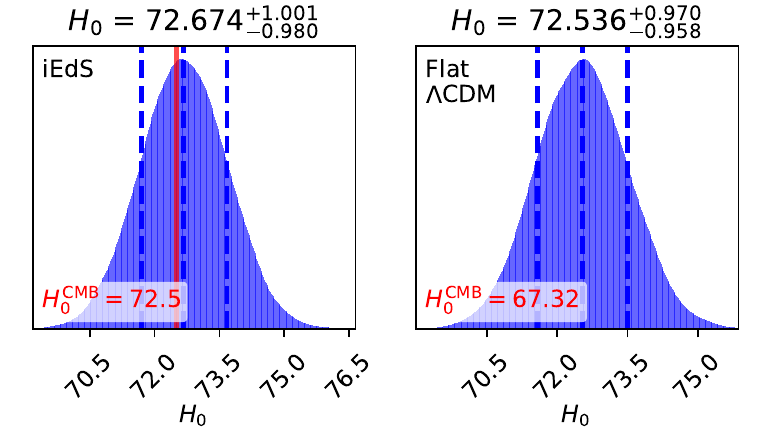}
     \caption{$H_0$ posterior distributions from Pantheon+ SNe Ia fits for the iEdS and flat $\Lambda$CDM models. Blue dashed lines show the posterior medians with 16th–84th percentile errors (given above each plot), and red lines (where visible) mark the reference $H_0^\mathrm{CMB}$ values (Table~\ref{tab:table1}). All $H_0$ values are in $\mathrm{km\ s^{-1}\ Mpc^{-1}}$. $H_0$ was fitted jointly with the nuisance parameters $\alpha$, $\beta$, $\gamma$, and $M_B$; the corresponding posterior distributions are available in our code repository~\cite{Zenodo_repo}.}
     \label{fig:fig4}
\end{figure}

\begin{figure}
 	\includegraphics[width=\columnwidth]{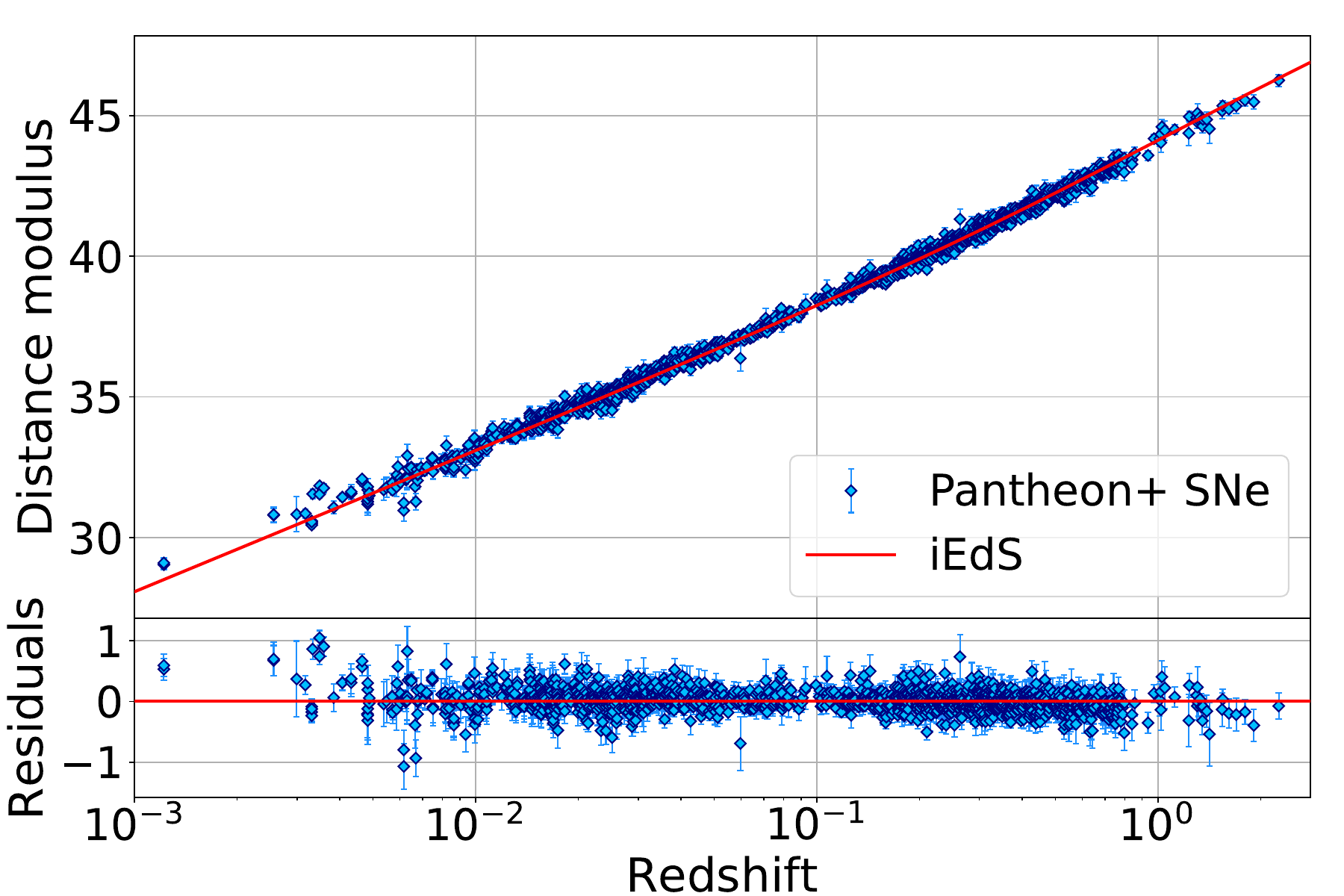}
     \caption{Distance moduli for $1701$ Pantheon+ SN Ia observations (shown at $z=z_\mathrm{HD}$) and for the best-fit iEdS model (see Table~\ref{tab:table2}). The lower panel shows the residuals relative to the model $\mu(z)$ curve.}
     \label{fig:fig5}
\end{figure}

\begin{figure}
 	\includegraphics[width=\columnwidth]{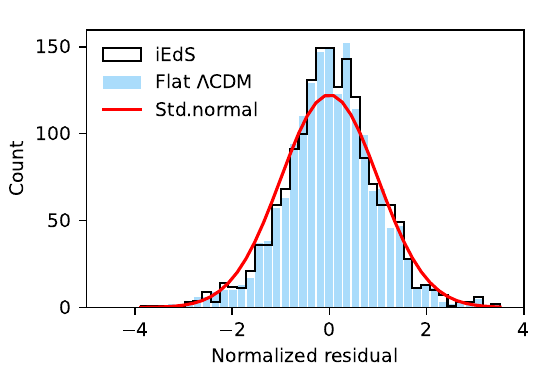}
     \caption{Histograms of standardized residuals (fit residuals normalized using $L$ from the $C=LL^T$ Cholesky factorization of the covariance) for the best-fit iEdS and flat $\Lambda$CDM models fitted to the Pantheon+ SNe. The same $>4\sigma$ outlier was excluded from the $1701$ data points in both fits. The red curve shows the standard normal distribution expected for the true model. Low-value residuals are overrepresented, with sample standard deviations of ${\sigma=0.95}$ for both the iEdS and $\Lambda$CDM fits.}
     \label{fig:fig6}
\end{figure}

Model evaluation, AD testing, and result presentation followed the same procedure as for the BAO analysis. The $H_0$ posteriors from the SN fits are shown in Fig.~\ref{fig:fig4}, full posterior corner plots are available in our code repository~\cite{Zenodo_repo}, and the results are listed in Table~\ref{tab:table2}. Fig.~\ref{fig:fig5} shows the best-fit iEdS $\mu(z)$ curve together with the 1701 Pantheon+ SN distance moduli. The $\chi^2$ values in Table~\ref{tab:table2} were computed using all SN data points except one that deviates from both models by more than $4\sigma$. Despite differing statistical performances, both models show systematic overfitting of the Pantheon+ sample (as illustrated in Fig.~\ref{fig:fig6}), with standardized residual dispersions of $\sigma=0.95$ for both iEdS and $\Lambda$CDM (compared to $\sigma=1$ for an adequate model), and neither satisfying the $p_\mathrm{AD}\geq 0.05$ consistency threshold. This suggests limitations in the Pantheon+ uncertainty estimates or in the model assumptions, consistent with earlier findings~\cite{Keeley_et_al_2024}. The iEdS fully resolves all $H_0$ tensions and yields a $p_\mathrm{AD}$ two times higher than that of $\Lambda$CDM, but at the cost of significantly higher $\chi^2$ and lower $\log_{10}\mathcal{B}$. These mixed outcomes underscore the need to reassess both theoretical and observational systematics before drawing firm model preferences from SNe Ia data.

\section{Conclusion}
\label{sec:Conclusion}

The iEdS model tested here is a specific realization of a broader cosmological framework which is based on the local-to-global approach defined by principles (i)–(ii) (see Sec.~\ref{sec:Theory}). In this framework, the accelerated expansion emerges as a purely gravitational effect within general relativity, without invoking alternative gravity theories or scalar fields as dark-energy components. Despite the analytic approximations in Eqs.~\eqref{eq:approx1}–\eqref{eq:approx2}, Sec.~\ref{sec:Data_fits} demonstrates that iEdS cosmologies can match $\Lambda$CDM in fitting CMB, BAO, and SNe Ia data, while resolving the Hubble tension (see Tables~\ref{tab:table1}–\ref{tab:table2}). The way forward is to develop a realistic model of structure formation—specifically for ${\sigma_H^2/H^2}$ or ${\left\langle H_i^2 \right\rangle/H^2}$—through simulations and observations, to construct the global iEdS evolution and test it against precision cosmological probes. As we demonstrated here, such a model could eliminate the need for dark energy, since $w_\mathrm{K}$ in Eq.~\eqref{eq:wK}, unlike $w_\mathrm{DE}$, is fully determined by matter and curvature.

The Planck 2018 baseline $\Lambda$CDM model gives an age of the universe of ${t_0=13.797\pm 0.023\ \mathrm{Gyr}}$. For the flat $\Lambda$CDM model fitted to Pantheon+ SNe~\cite{Brout_et_al_2022b}, the derived age is ${t_0=12.4\pm 0.3\ \mathrm{Gyr}}$, in ${\sim 3\sigma}$ tension with the globular-cluster estimate ${t_0=13.6\pm 0.3\ \mathrm{Gyr}}$~\cite{Valcin_et_al_2025}. Our \texttt{camb} run yield ${t_0=13.669\ \mathrm{Gyr}}$ for iEdS, consistent with globular-cluster ages. The corresponding ${S_8=0.822}$ value is slightly lower but consistent with the Planck $\Lambda$CDM result ${S_8=0.834\pm 0.016}$~\cite{Planck_2018}. However, without a detailed theory of structure formation and dedicated $S_8$ fits to late-time structure data within the iEdS framework, no conclusions can yet be drawn about how iEdS cosmology affects the $S_8$ tension.

Similarly to a $\Lambda$CDM universe, an iEdS one avoids both recollapse (Big Crunch) and divergent expansion (Big Rip), asymptotically approaching a dark, thermodynamically frozen equilibrium state known as the Big Freeze. In contrast, in a universe undergoing eternal accelerated expansion, such as that described by the $\Lambda$CDM model, the maximum comoving distances that can be reached or observed are finite. This is not the case in the iEdS model, whose dynamics asymptotically approach linear expansion, $a(t)\propto t$. If our universe indeed follows an iEdS evolution, both comoving horizons are infinite, revealing a cosmos without unreachable or unobservable realms, where every part of the universe is, in principle, open to discovery, regardless of whether the universe is finite or infinite in size.

\begin{acknowledgments}
The authors would like to thank Bence B\'ecsy, Franciska M. Constans, and Attila Cs\'ot\'o for fruitful discussions during the project. PR thanks Adam Riess for guiding advice that helped shape this work. This project has received funding from the HUN-REN Hungarian Research Network and was supported by the NKFIH excellence grant TKP2021-NKTA-64.
\end{acknowledgments}

\appendix

\section{The Expansion-Rate Variance}
\label{app:sigma_H}

We derive the leading-order dependence of ${\sigma_H^2 \equiv \left\langle H_i^2 \right\rangle-\left\langle H_i \right\rangle^2}$ on local matter and curvature fluctuations, and clarify how our local-to-global framework differs from standard global-to-local cosmology and cosmic-backreaction approaches.

From Eq.~\eqref{eq:local_Friedmann}, we have
\begin{equation}\label{eq:sqrt_Hi}
H_i=\pm \sqrt{\left\langle H_i^2 \right\rangle} \sqrt{1+\widetilde{\Omega}_{\mathrm{m},i}+\widetilde{\Omega}_{\mathrm{k},i}} ,
\end{equation}
where ${\widetilde{\Omega}_{\mathrm{m},i}\equiv \Omega_{\mathrm{m},i}-\left\langle \Omega_{\mathrm{m},i} \right\rangle}$ and ${\widetilde{\Omega}_{\mathrm{k},i}\equiv \Omega_{\mathrm{k},i}-\left\langle \Omega_{\mathrm{k},i} \right\rangle}$, with $\langle \widetilde{\Omega}_{\mathrm{m},i}\rangle = \langle \widetilde{\Omega}_{\mathrm{k},i}\rangle = 0$ and $\left\langle \Omega_{\mathrm{m},i} \right\rangle +\left\langle \Omega_{\mathrm{k},i} \right\rangle = 1$.

For ${\widetilde{\Omega}_{\mathrm{m},i},\widetilde{\Omega}_{\mathrm{k},i}\ll 1}$, Eq.~\eqref{eq:sqrt_Hi} yields
\begin{equation}\label{eq:sqrt_Hi_approx}
H_i\approx \pm \sqrt{\left\langle H_i^2 \right\rangle} \left( 1+\frac{1}{2}\widetilde{\Omega}_{\mathrm{m},i}+\frac{1}{2}\widetilde{\Omega}_{\mathrm{k},i} \right) ,
\end{equation}
implying $\left\langle H_i \right\rangle\approx \pm \sqrt{\left\langle H_i^2 \right\rangle}$. The leading-order squared deviation is then
\begin{equation}\label{eq:de_Hisq_approx}
\left( H_i - \left\langle H_i \right\rangle \right)^2 = \frac{\left\langle H_i^2 \right\rangle}{4} \left(\widetilde{\Omega}_{\mathrm{m},i}^2 + \widetilde{\Omega}_{\mathrm{k},i}^2 + 2 \widetilde{\Omega}_{\mathrm{m},i} \widetilde{\Omega}_{\mathrm{k},i}\right).
\end{equation}

Taking the volume average gives
\begin{equation}\label{eq:sigmaH_approx}
\frac{\sigma_H^2}{\left\langle H_i^2 \right\rangle} = \frac{1}{4} \left(\sigma_\mathrm{m}^2 + \sigma_\mathrm{k}^2 + 2 \rho_\mathrm{mk}\sigma_\mathrm{m}\sigma_\mathrm{k}\right),
\end{equation}
where ${\sigma_\mathrm{m}^2=\langle\widetilde{\Omega}_{\mathrm{m},i}^2\rangle}$, ${\sigma_\mathrm{k}^2=\langle\widetilde{\Omega}_{\mathrm{k},i}^2\rangle}$, and ${\rho_\mathrm{mk}\equiv \langle \widetilde{\Omega}_{\mathrm{m},i} \widetilde{\Omega}_{\mathrm{k},i}\rangle \sigma_\mathrm{m}^{-1} \sigma_\mathrm{k}^{-1}}$ is the Pearson correlation coefficient between $\widetilde{\Omega}_{\mathrm{m},i}$ and $\widetilde{\Omega}_{\mathrm{k},i}$.

From principle (ii) in Sec.~\ref{sec:Theory}, $\widetilde{\Omega}_{\mathrm{m},i}$ and $\widetilde{\Omega}_{\mathrm{k},i}$ are uncorrelated (${\rho_\mathrm{mk}=0}$), yielding
\begin{equation}\label{eq:sigma_H_iEdS}
\left[ \frac{\sigma_H^2}{\left\langle H_i^2 \right\rangle} \right]_\mathrm{iEdS} = \frac{\sigma_\mathrm{m}^2 + \sigma_\mathrm{k}^2}{4}.
\end{equation}
In homogeneous cosmology, ${\sigma_\mathrm{m}=\sigma_\mathrm{k}=0}$, while in cosmic-backreaction studies ${\sigma_\mathrm{m}=\sigma_\mathrm{k}>0}$ with ${\rho_\mathrm{mk}=-1}$, resulting in $\sigma_H=0$ in both cases.

\section{Impact of the LRG1 BAO Data Points}
\label{app:BAO_full_fit}

Table~\ref{tab:BAO_full_fit} summarizes the results of our DESI DR2 BAO fits including the LRG1 ${D_\mathrm{M}/r_\mathrm{d}}$ and ${D_\mathrm{H}/r_\mathrm{d}}$ measurements at ${z=0.51}$. The listed quantities are the same as in Table~\ref{tab:table2} (see Sec.~\ref{sec:BAO}). We publish the corresponding $H_0$ posterior in our public repository~\cite{Zenodo_repo}.

While both models remain consistent with the BAO data, inclusion of the LRG1 datapoints degrades the fits and decisively favors $\Lambda$CDM. Nevertheless, the best-fit $H_0$ in the iEdS model moves closer to the CMB and SN Ia determinations, whereas the Hubble tension in $\Lambda$CDM is strengthened.

\begin{table}
\caption{\label{tab:BAO_full_fit} Model comparison results from DESI DR2 BAO fits including the LRG1 data points.}
\renewcommand{\arraystretch}{1.25}
\begin{ruledtabular}
\begin{tabular}{lcc}
& iEdS & $\Lambda$CDM \\
\colrule
\colrule
$p_\mathrm{AD}\ [10^{-2}]$ & $37.11$ & $76.0$ \\
$\chi^2$ & $23.168$ & $10.436$ \\
$\log_{10}\mathcal{B}$ & $-3.789$ & $0$ \\
$H_0\ [\mathrm{km}\ \mathrm{s}^{-1}\ \mathrm{Mpc}^{-1}]$ & $72.47\pm 0.33$ & $68.90\pm 0.36$ \\
$\left| \Delta H_0^\mathrm{CMB} \right|\ [\sigma]$ & $\color{Green} 0.10\ \color{black}(\color{Green}0.06\color{black})$ & $\color{red} 4.35\ \color{black}(\color{Green}2.43\color{black})$ \\
$\left| \Delta H_0^\mathrm{SN} \right|\ [\sigma]$ & $\color{Green} 0.20$ & $\color{red} 3.51$ \\
\end{tabular}
\end{ruledtabular}
\end{table}

\bibliography{iEdS}

\end{document}